\documentclass{emulateapj}
\usepackage{apjfonts}
\begin{document}

\title{SN~2005cg: Explosion Physics and Circumstellar Interaction of a Normal Type Ia Supernova in a Low-Luminosity Host\footnotemark[1]}

\author{
  Robert Quimby\altaffilmark{2},
  Peter H\"{o}flich\altaffilmark{2,3},
  Sheila J. Kannappan\altaffilmark{2},
  Eli Rykoff\altaffilmark{4},
  Wiphu Rujopakarn\altaffilmark{4},
  Carl W. Akerlof\altaffilmark{4},
  Christopher L. Gerardy\altaffilmark{5},
  J. Craig Wheeler\altaffilmark{2}
}
\footnotetext[1]{Based on observations obtained with the Hobby-Eberly
Telescope, which is a joint project of the University of Texas at
Austin, the Pennsylvania State University, Stanford University,
Ludwig-Maximilians-Universit\"{a}t M\"{u}nchen, and
Georg-August-Universit\"{a}t G\"{o}ttingen.}
\altaffiltext{2}{Department of Astronomy, University of Texas, Austin, TX 78712, USA}
\altaffiltext{3}{McDonald Observatory, University of Texas, Austin, TX 78712, USA}
\altaffiltext{4}{Randall Laboratory, University of Michigan, Ann Arbor, MI 48104-112, USA}
\altaffiltext{5}{Astrophysics Group, Blackett Laboratory, Imperial College, London, UK, SW7 2AZ}

\begin{abstract}
We present the spectral evolution, light curve, and corresponding
interpretation for the ``normal-bright'' Type Ia Supernova 2005cg
discovered by ROTSE-IIIc. The host is a low-luminosity
($\rm{M}_{r}=-16.75$), blue galaxy with strong indications of active
star formation and an environment similar to that expected for SNe~Ia
at high redshifts. Early-time ($t\sim -10$ days) optical spectra
obtained with the HET reveal an asymmetric, triangular-shaped
\ion{Si}{2} absorption feature at about 6100 \AA\ with a sharp
transition to the continuum at a blue shift of about 24,000 km
s$^{-1}$. By 4 days before maximum, the \ion{Si}{2} absorption feature
becomes symmetric with smoothly curved sides. Similar \ion{Si}{2}
profile evolution has previously been observed in other supernovae,
and is predicted by some explosion models, but its significance has
not been fully recognized. Although the spectra predicted by pure
deflagration and delayed detonation models are similar near maximum
light, they predict qualitatively different chemical abundances in the
outer layers and thus give qualitatively different spectra at the
earliest phases. The Si line observed in SN~2005cg at early times
requires the presence of burning products at high velocities and the
triangular shape is likely to be formed in an extended region of
slowly declining Si abundance that characterizes delayed detonation
models. The spectra show a high-velocity \ion{Ca}{2} IR feature that
coincides in velocity space with the \ion{Si}{2} cutoff. This supports
the interpretation that the \ion{Ca}{2} is formed when the outer
layers of the SN ejecta sweep up about $5 \times 10^{-3} M_\odot$ of
material within the progenitor system.  We compare our results with
other ``Branch-normal'' SNe~Ia with early time spectra, namely
SN~2003du, 1999ee and 1994D.  Although the expansion velocities based
on their \ion{Si}{2} absorption minima differ, all show
triangular-shaped profiles and velocity cutoffs between 23,000 and
25,000 km s$^{-1}$, which are consistent with the Doppler shifts of
their respective high-velocity \ion{Ca}{2} IR features. SN~1990N-like
objects, however, showed distinctly different behavior that may
suggest separate progenitor sub-classes.

\end{abstract}


\section{Introduction}

Type Ia supernovae (SNe~Ia) are important phenomena in shaping the
metal enrichment history of the Universe, excellent tools for probing
its expansion history \citep{riess1998,perlmutter1999}, and they also
provide a unique laboratory to study combustion physics,
hydrodynamics, plus nuclear and atomic processes. SN~2005cg can shed
light on a number of questions involving the physics of supernovae and their
use in cosmology. The host is a low-luminosity galaxy, raising the
possibility of a low metallicity for the SN progenitor, and the
galaxy colors suggest a substantial population of young stars. Line
profiles of \ion{Si}{2} constrain the chemical structure of the outer
layers and provide a test for nuclear burning models.

The most favored models for SNe~Ia involve a white dwarf (WD) near the
Chandrasekhar mass accreting matter from a binary companion, which
eventually results in a thermonuclear explosion. One of the key
questions of physics is how the burning front propagates: whether it
remains a subsonic deflagration front, or turns into a weak detonation
as in the so-called delayed detonation (DD) models. A key difference
between these possibilities is the velocity range of the explosion
products. In deflagration models the outer layers of the envelope
expand with velocities close to the sound speed
\citep{hoflich_khokhlov1996} and, by causality, the subsonic
deflagration cannot keep up. Thus the outer layers remain unburned
C/O. In contrast, nearly the entire WD undergoes burning in delayed
detonation models, at least for normal-bright SNe~Ia. The structures
of the classical deflagration model W7 and delayed detonation models
are very similar inside the region of incomplete Si burning where the
spectra at maximum light are formed, but they are very different in
the outer layers responsible for line formation at very early
times. For recent reviews on SNe~Ia, see
\citet{branch1998,hoflich2005}.

High-velocity \ion{Ca}{2} (\ion{Ca}{2} HV) has been found to be a
common feature in SNe~Ia (e.g.  SN~1994D;
\citealt{hatano1999,fisher2000}; SN~1999ee; \citealt{mazzali2005a};
SN~2000cx; \citealt{li2001,thomas2004}; SN~2001el; \citealt{wang2003};
SN~2003du; \citealt{gerardy2004}). \citet{wang2003} showed that this
feature in SN~2001el was kinematically distinct from the photospheric
\ion{Ca}{2} IR triplet and suggested it could be a consequence of
nuclear burning in the WD (perhaps during the deflagration to
detonation transition), which causes the ejection of a high-velocity,
Ca-rich filament, or that it might be attributed to the surrounding
accretion disk, which has undergone nuclear burning to increase the Ca
abundance.  \citet{gerardy2004} studied the formation of the
\ion{Ca}{2} HV feature based on detailed NLTE-models that included
interaction with circumstellar material (CSM). They showed that the
\ion{Ca}{2} HV feature and its evolution with time could be understood
in the framework of the interaction of the ejecta with a circumstellar
shell of solar composition. They predicted a corresponding blue cutoff
in the \ion{Si}{2} absorption feature that should be visible at early
times.

In this paper we discuss SN~2005cg, a normal SN~Ia discovered by the
wide field ROTSE-III sky patrol search \citep{rykoff2005a}. In
\S{\ref{obs}} we give the early broad-band ROTSE-IIIc light curve and
spectral evolution recorded by the Hobby-Eberly Telescope (HET). We
discuss constraints on explosion models from the \ion{Si}{2} line
profile in \S{\ref{SiII}} and give an interpretation of the
\ion{Ca}{2} HV feature in \S{\ref{csm}}. In \S{\ref{host}} we discuss
properties of the host galaxy and its implications for the
progenitor. In \S{\ref{other}}, SN~2005cg is put into context with
other SNe~Ia.  Conclusions and discussion are presented in
\S{\ref{conclusions}}.

\section{Observations}\label{obs}

\begin{figure}
\epsscale{1.25}
\plotone{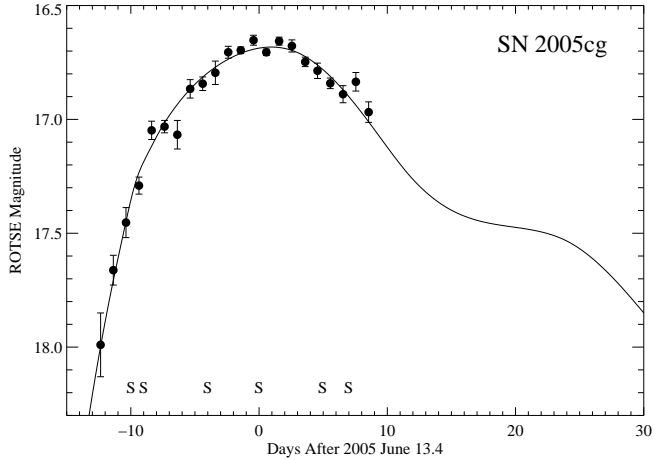}
\caption{ROTSE-IIIc unfiltered light curve of SN~2005cg (magnitudes
calibrated against the Sloan r-band). The curve is the fitted R-band
template from \citet{knop2003} dilated in time by $(1+z)/s$ where
$s=1.07\pm0.02$ is the stretch parameter. Spectral epochs are marked
with an ``S''.}
\label{lc}
\end{figure}

SN~2005cg was discovered on 2005 June 1 (UT) at about 18.0 magnitude
using the 0.45 m ROTSE-IIIc telescope at the High-Energy Stereoscopic
Systems site in Namibia \citep{rykoff2005b}. The supernova is located
at $\alpha=21h10m50.42s$, $\delta=+00^{\circ}12'07.6''$. Daily
follow-up observations with ROTSE-IIIc since discovery give the
unfiltered broad-band light curve shown in Figure~\ref{lc}. We
processed the data with a customized version of the DAOPHOT
PSF-fitting package \citep{stetson87} ported to IDL by
\citet{landsman89}. The magnitude zero-point for each image was
calculated from the median offset of fiducial reference stars to the
SDSS $r$-band values \citep{aaaaa05}. To determine the date of maximum
light, we fit the SN~Ia R-band template of \citet{knop2003} to our
data, where the phases are relative to the B-band maximum. The best
fit puts the maximum on June 13.4 with a formal error of $\pm$0.1
days, and shows that SN~2005cg is a normal SN~Ia.

\begin{figure}
\epsscale{1.25}
\plotone{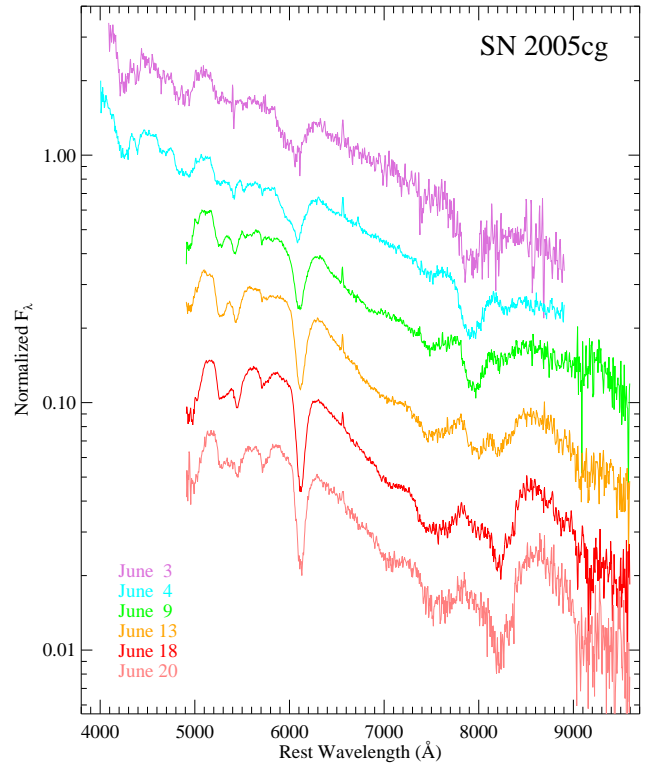}
\caption{HET/LRS spectral evolution of SN~2005cg from -10 to +7 days
relative to the derived June 13th maximum. The spectra were normalized
between 6900 and 7100\AA\ and shifted by factors of 2 for clarity.}
\label{spec}
\end{figure}

We obtained a low resolution (R$\sim$300) optical spectrum using the
Marcario Low Resolution Spectrograph (LRS; \citealt{hill1998}) on the
Hobby-Eberly Telescope (HET; \citealt{ramsey1998}) on 2005 June 3,
which showed SN~2005cg to be a Type Ia supernova
\citep{quimby2005}. The effective wavelength range is 4100-7800\AA\
for these data and for a second HET/LRS spectrum taken on 2005 June 4;
at longer wavelengths order-overlap begins to contaminate the
spectrum. For subsequent spectral observations we used an OG515
blocking filter, giving an effective coverage of 5150-10000\AA. We
used the standard star Wolf1346
\citep{massey1988,massey_gronwall1990}, observed with both setups on
2005 June 4, to perform relative spectrophotometric calibration and to
correct for telluric absorption. The spectral evolution of SN~2005cg
is presented in Figure~\ref{spec}.

The host galaxy of SN~2005cg was observed prior to the explosion by
the Sloan Digital Sky survey and designated SDSS J211050.45+001206.7
\citep{aaaaa05}. The host has a $g$ magnitude of $19.719 \pm 0.021$
and a Petrosian radius of $1.991''$. Identifying narrow emission
features in the HET spectra at 6768, 5013, and 5165\AA\ as H-alpha,
H-beta, and [\ion{O}{3}] 5007\AA, respectively, we derive a redshift
of $z=0.0313 \pm 0.0010$.

\section{The \ion{Si}{2} line profiles}\label{SiII}

\begin{figure*}
\includegraphics[width=8.5cm,angle=270]{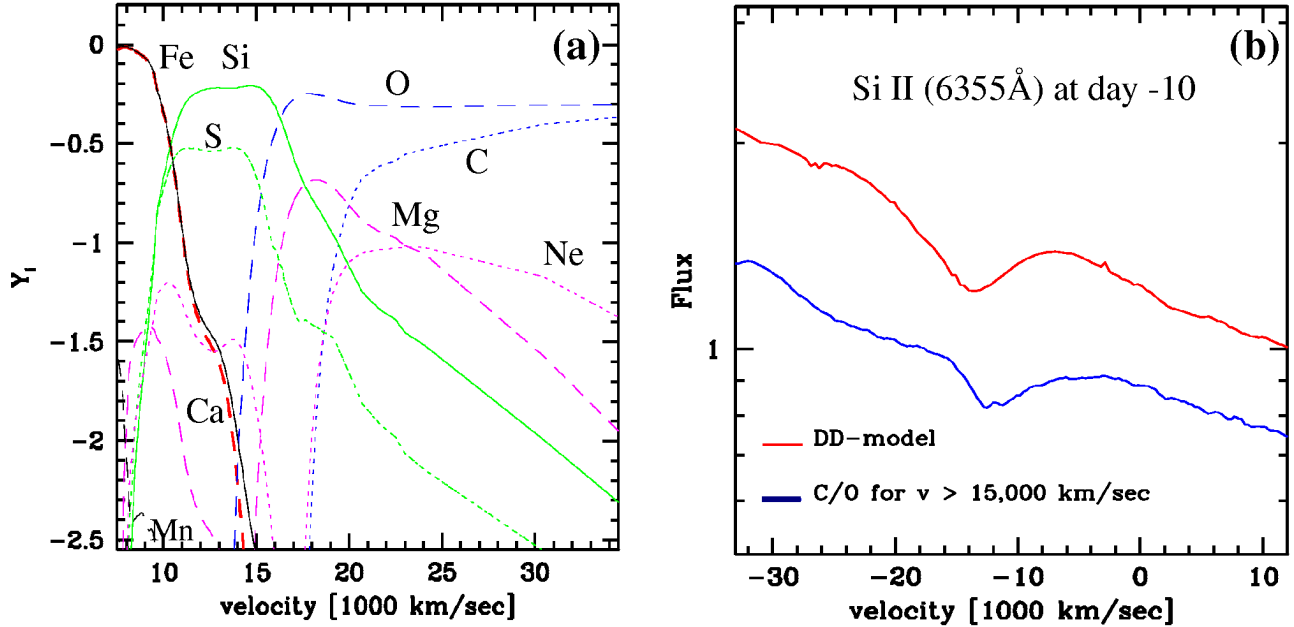}
\caption{(a) Abundance structure of the delayed detonation model
$5p0z22.25$ for a normal-bright SNe~Ia \citep{hoflich2002} and (b)
early time model spectra of the \ion{Si}{2} 6355 \AA\ line at about
8.5 days after the explosion.  Between 11,000 and 15,000 km s$^{-1}$,
products of explosive O burning have almost constant abundance ratios
because the isotopic distribution is given by quasi-nuclear
equilibrium within the Si-group.  Layers with $\rm{v} > 15,000$ km
s$^{-1}$ undergo explosive carbon burning, but about 10 and 1 \% Si is
freshly synthesized up to velocities of 19,000 and 28,000 km s$^{-1}$,
respectively.  In the early DD model spectra (top red line), the
\ion{Si}{2} absorption feature shows a triangular-shape with an
extended blue wing and a small blend due to \ion{Fe}{2} at about
26,000 km s$^{-1}$. For comparison, we also present a spectrum (bottom
blue line) assuming unburned C/O matter for $\rm{v} > 15,000$ km
s$^{-1}$ similar to W7 (see text).}
\label{model}
\end{figure*}

The first two spectra of SN~2005cg obtained about 9 and 10 days before
maximum, respectively, reveal an asymmetric ``triangular-shaped''
\ion{Si}{2} absorption profile characterized by an approximately
linear (in log flux) slope of the line wings and an extended blue
wing. There is also a sharp break in slope or ``cutoff'' between the
blue wing of the \ion{Si}{2} line and the continuum. In subsequent
spectra (4 days before maximum and later) the profile becomes more
curved and symmetric. Although this time dependence has been observed
previously (see \S{\ref{other}}) and indeed is predicted by delayed
detonation models \citep{hoflich1995}, the implications have mostly
been ignored. In SN~2005cg, the blue wing of the \ion{Si}{2} in the
two earliest spectra extends out to 24,000 to 25,000 km s$^{-1}$; this
requires a significant Si abundance in the outer most layers.  Thus,
the Si abundance must not decline too rapidly with radius and hence
velocity. The DD models depicted in Figure~\ref{model} naturally give
an extended distribution of Si and line profiles that approximate the
early observations. For comparison, we also give the line profile
assuming that no burning took place above velocities of 15,000 km
s$^{-1}$ corresponding to the deflagration model W7
\citep{nomoto1984}; this profile lacks the extended blue Si wing.
Outward mixing of Si has been suggested as a means to provide higher
velocity \ion{Si}{2} \citep{harkness1986}. In principle, mixing
properties of deflagration models may be tuned to reproduce chemical
profiles similar to DD models; however, the composition of
high-velocity mixed matter would consist of explosive oxygen burning
products, namely Si and S, and lack explosive carbon burning products,
such as Mg. Typically, SNe~Ia show strong \ion{Mg}{2} lines in the
near IR at high velocities that require a Mg abundance of a few
percent, consistent with layers undergoing explosive carbon burning
(e.g. \citealt{bowers1997,wheeler1998,hoflich2002, marion2003}).

In models, the triangular shape of the \ion{Si}{2} profile requires
the line formation to take place in a region of Si concentration that
declines faster with distance than the density slope. The profile is
linear in log flux if the abundance logarithm declines linearly with
velocity, and its extent in velocity space is linked to the abundance
if the effective optical depth is small. At very early times when the
effective optical depth is large close to the absorption trough, the
linear profile would be to the blue of a ``round absorption core.''
At later times the highest velocity material has thinned, thus the
model spectra show deeper layers of the explosion and probe a more
restricted region of velocity space due to the more shallow velocity
gradient. During these phases, the classic ``6150'' \ion{Si}{2}
profile is seen with a gentle roll over on the blue wing as it meets
with the continuum.

\section{Evidence for CSM Interaction}\label{csm}

The early spectra of SN~2005cg show strong absorption around 7900\AA,
which we attribute to high velocity \ion{Ca}{2} (8498, 8542,
8662\AA). This feature is common to most, if not all, SNe~Ia observed
before maximum light in the appropriate wavelength range
\citep{mazzali2005b}.  \citet{gerardy2004} credit this absorption to a
surrounding ($< 6 \times 10^{14}$cm) region of hydrogen-rich material,
possibly from an accretion disk, Roche lobe, or common envelope, swept
up by the outer layers of the SN ejecta. \citet{mattila2005} find an
upper limit of 0.03 M$_\odot$ for solar abundance material present
within the SN~2001el explosion site, which is consistent with the 0.02
M$_\odot$ derived by \citet{gerardy2004} for SN~2003du. Both papers
rule out a continuous wind extending beyond $\sim 10^{15}$cm.

\begin{figure}
\epsscale{1.25}
\plotone{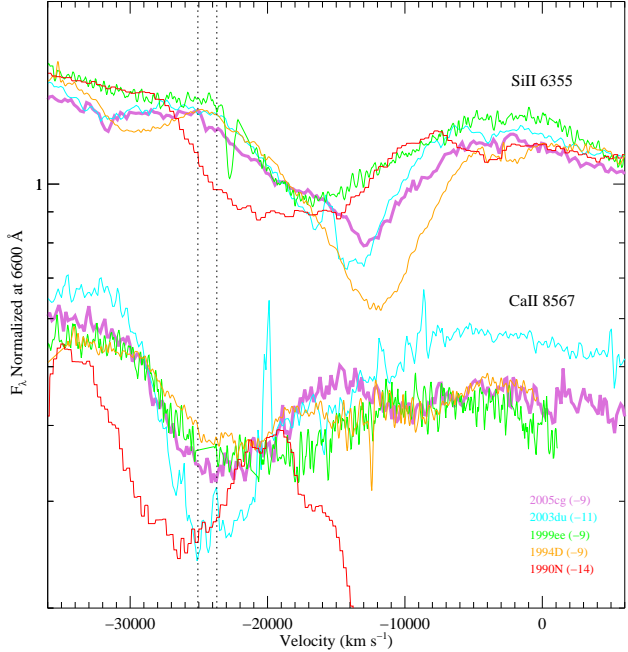}
\caption{\ion{Si}{2} and \ion{Ca}{2} line velocities for normal-bright
SNe~Ia at early epochs. The vertical dashed lines mark the range of
\ion{Si}{2} cutoff velocities and are seen to align with the blue edge
of the \ion{Ca}{2} HV absorption minima as expected for an optically
thin shell.}
\label{line_vel}
\end{figure}

 Since burning a given mass of C/O to Si or to heavier elements
releases roughly the same amount of energy, the velocity of the outer
layers of the SN ejecta will remain approximately constant for a
variety of explosion models provided the Chandrasekhar mass WD is
completely burned. Following \citet{gerardy2004}, we can therefore
determine the mass of circumstellar material (CSM) swept into the
shell simply by requiring momentum conservation. With this and a
polytropic density gradient for the WD, we can then find
$M_{\rm{CSM}}$ by measuring the shell velocity from the \ion{Ca}{2} HV
line profile. Figure~\ref{line_vel} shows the \ion{Si}{2} and
\ion{Ca}{2} line profiles from the June 4 spectrum of SN~2005cg
plotted in velocity space relative to 6355\AA\ and 8567\AA,
respectively. These wavelengths represent the average of the doublet
and triplet lines respectively weighted by their $gf$ values. The
shell velocity is $\sim$23,000 km s$^{-1}$ measured from the minimum
of the \ion{Ca}{2} HV, giving 5 to 7 $\times 10^{-3}$ M$_\odot$ of the
CSM accumulated in the shell.

The \ion{Si}{2} absorption profile can be understood in the framework
of DD models (see \S{\ref{SiII}}) but neglecting CSM interaction the wings
are expected to extend to $\approx 30,000$ km s$^{-1}$
(Figure~\ref{model}). With an interacting shell, the outer SN ejecta
are slowed down. For SN~2005cg, the blue wing of the absorption
terminates abruptly at $\sim$24,000 km s$^{-1}$
(Figure~\ref{line_vel}). We interpret this cutoff as a truncation in
the velocity distribution created when the outer ejecta layers are
decelerated as they interact with the CSM.

The changing optical depth of the thinning shell will move the
apparent \ion{Ca}{2} HV minimum to the red with time. Even so, the
coincidence of the Si cutoff with the blue edge of the \ion{Ca}{2} HV
minimum seen in Figure~\ref{line_vel} supports the interpretation
that the Si cutoff is caused by the CSM shell . We note there is a
slight dip in the \ion{Si}{2} line profile near the cutoff
velocity. While this may simply be due to \ion{Fe}{2} absorption, it
is possible this is a signature of the Si produced in the outer layers
of the explosion piling up in the shell.

\section{Host Galaxy and Progenitor Constraints}\label{host}

\begin{figure}
\epsscale{1.25}
\plotone{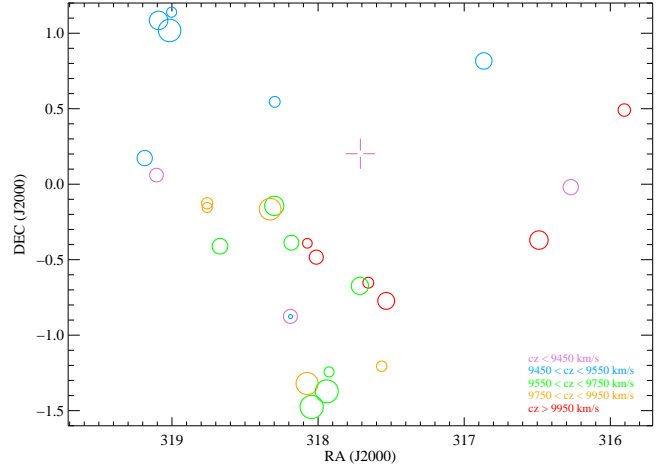}
\caption{Galaxies within $\pm$1000 km s$^{-1}$ of the SN~2005cg host
galaxy (cross).  One degree is $\sim$2.5 Mpc at this distance.  No
close companion with measured redshift is known, but the global
environment is overdense. The elongated structure and low velocity
dispersion ($\sim$285 km s$^{-1}$) of the galaxy distribution imply a
filament or merging group environment rather than a cluster.  Visual
inspection of galaxy morphologies confirms a field mix of early and
late types. The size of the symbols represents their relative absolute
magnitudes, while the color indicates the redshift.  }
\label{cluster}
\end{figure}

The host of SN~2005cg is a low-mass dwarf galaxy in a filament or
merging group environment (Figure~\ref{cluster}), raising some
interesting questions about SNe~Ia progenitors and the properties of
high-$z$ SNe~Ia.  Using the calibrations of \citet{kannappan2004} and
references therein, the host galaxy has $u-r$ color 1.27, luminosity
$M_r=-16.75$, half-light radius $r_{50} ^r$ $\sim$ 0.6 kpc, stellar
mass $\sim$ 4$\times$$10^8$ M$_\odot$, and atomic-gas--to--stellar
mass ratio $\sim$ 1:1 (with factor of 2-3 uncertainty on the latter
two). From the Tully-Fisher relation in $r$, the galaxy's internal
velocity is $\ga50$ km s$^{-1}$, implying a binding energy high enough
to avoid gas blow-away \citep{maclow_ferrara1999}. Nearby analogues
with similar colors, masses, and environments display knotty irregular
morphology or post-interaction distortions (based on the Nearby Field
Galaxy Survey, \citealt{jansen2000}; environments computed following
\citealt{grogin_geller1998}). Among analogues with similarly compact
radii, most systems have blue-centered color gradients and strong star
formation (EW(H$\alpha$) $\sim$ 25--200), suggesting starburst
activity. Gas metallicities are subsolar (log(O/H) + 12 = 8.5--8.8)
and 21-cm data yield high gas-to-stellar mass ratios $\sim$0.5--1.  We
infer that the dwarf host of SN~2005cg is likely to have many
similarities to high-$z$ SN hosts: low mass, high gas content, strong
star formation, low metallicity, and perhaps environmental
perturbations.

While we cannot directly constrain the age of the SN progenitor, a
young progenitor is an intriguing
possibility \citep{oemler_tinsley1979}. The $ugriz$ photometry of the
host galaxy suggests at least two stellar populations, with fits to
the models of \citet{bruzual_charlot2003} generally favoring a
10--20\% young, solar metallicity population (25--100 Myr) mixed with
a larger intermediate-age metal-poor population (640--900 Myr).
However, an underlying 2.5--5 Gyr population is not ruled out.  In
fact, most resolved studies of nearby dwarf galaxies have found traces
of ancient stellar populations (e.g. \citealt{grebel_gallagher2004}),
consistent with a long and bursty star formation history.  For the
above-mentioned analogues of our SN host, nominal gas consumption
timescales range from 2-13 Gyr, but continuing gas accretion is
likely.

\section{SN~2005cg in Context With Other SNe~Ia}\label{other}

A search through the
SUSPECT\footnote{\url{http://bruford.nhn.ou.edu/~suspect/index1.html}}
spectral archive reveals that the triangular \ion{Si}{2} and the
\ion{Ca}{2} HV features are common to other normal SNe~Ia around 10
days before maximum light. Included in Figure~\ref{line_vel} are SN
2003du at -11 days \citep{anupama2005}, SN~1999ee at -9 days
\citep{hamuy2002}, and SN~1994D at -9 days \citep{patat1996}. As with
SN~2005cg, all show a remarkably similar blue \ion{Si}{2} wing up to
$\sim 24,000$ km s$^{-1}$ with corresponding \ion{Ca}{2} HV. Within
our interpretation, the triangular \ion{Si}{2} profiles suggest that
all these SN originate from similar explosion conditions. This
conclusion is also supported by the recent analysis of photospheric
expansion velocities ($\rm{v}_{ph}$) of \citet{benetti2005}. However,
this does not mean that all normal-bright SNe~Ia are the same.  In our
sample, SN 1990N at -14 days shows rather ``round'' \ion{Si}{2}
profiles \citep{leibundgut1991}. Moreover, SN~1990N shows very high
but rapidly declining $\rm{v}_{ph}$ up to a few days before maximum
light, followed by an almost constant $\rm{v}_{ph}$
\citep{benetti2005}. Both behaviors may be understood in the framework
of models that produce shell-like structures as expected in mergers or
pulsating delayed detonation models
\citep{khokhlov1993,hoflich_khokhlov1996}.

\section{Discussion and Conclusions}\label{conclusions}

SN~2005cg appears to be a Branch-normal SN~Ia in terms of its
light curve and spectra. It is therefore remarkable that this normal
SN~Ia may shed new light on thermonuclear explosions and SN research
in general.

Rare for low-z SNe~Ia, the host is a faint, blue galaxy with strong
indications for active star formation and an environment comparable to
SNe~Ia at high redshifts. Because low mass galaxies tend to be low in
metallicity, SN~2005cg is a good candidate to have a low metallicity
or young progenitor. As discussed in \S{\ref{host}}, the current
results are inconclusive, and we may never know whether SN~2005cg
originates from a young or old population. However, after the SN
fades, detailed analysis of the host galaxy should allow us to address
the question of metallicity. If it is low and homogeneous over the
galaxy, SN~2005cg may be the new empirical standard for a low
metallicity SNe~Ia.
 
The second remarkable feature is the method of discovery. SN~2005cg
was discovered in a blind, wide-field transient search by ROTSE-III in
a galaxy that would not have been targeted in traditional low redshift
searches.  Previously, there has been a difference in the way low and
high redshift SNe are sampled: high-z SNe are discovered in blind
searches while low-z SNe are found with targeted surveys biased to the
larger, more productive host galaxies. Recently, blind wide field
searches have been conducted to discover nearby SNe in a manner
consistent with the high-z searches (c.f. SNFactory,
\citealt{aldering2002}), netting several SNe~Ia in low-luminosity
hosts including SN~1999aw with its exceptionally faint
($\rm{M}_{\rm{B}} \sim -12$) host \citep{strolger2002}. These SNe~Ia
in low luminosity hosts deserve much closer scrutiny.

SN~2005cg draws attention to the importance of line profiles as
diagnostic tools. In the early spectra, the \ion{Si}{2} 6355\AA\ line
rises slowly up from the minimum to the blue edge where it sharply
meets with the continuum between 24,000 and 25,000 km s$^{-1}$. This
characteristic is commensurate with DD models that give a slow
decrease in \ion{Si}{2} production out to the edge of the WD; however,
it is inconsistent with pure deflagration models such as W7 that do
not burn the outer layers completely and hence do not predict
\ion{Si}{2} at velocities above $\sim 14,500$ km s$^{-1}$. In
principle, mixing may produce extended Si structures, but this would
be inconsistent with \ion{Mg}{2} features commonly seen in the IR
spectra.  SN~2005cg provides additional evidence that we need a
detonation phase in SNe~Ia explosions.

Another remarkable feature of SN~2005cg is the presence of a
high-velocity component in \ion{Ca}{2}. This high-velocity feature can
be well understood in the framework of an interaction of the rapidly
expanding ejecta with H or He-rich surroundings within the progenitor
system \citep{gerardy2004}. As predicted, the \ion{Ca}{2} HV coincides
with the high-velocity cutoff in the \ion{Si}{2} line. Although this
observation supports the model presented by \citet{gerardy2004},
circumstellar interaction is not necessarily a unique interpretation.

Finally, we compared SN~2005cg in the context of other Branch-normal
supernovae, and found them strikingly homogeneous in appearance both
with respect to the triangular \ion{Si}{2} profiles and in the
high-velocity cutoffs in Si, which are consistent with the \ion{Ca}{2}
HV features.  Combining our sample with that of Gerardy et al. (2004),
Doppler shifts indicate CSM shell masses in the range between $5$ and
$ 40 \times 10^{-3} ~M_\odot$.  Alternative interpretations of the
high-velocity \ion{Ca}{2} (see introduction) seem to be less
satisfactory.  The long-lived nature disfavors interpreting the
feature as \ion{Ca}{3} recombining to \ion{Ca}{2} \citep{hoflich1998},
while the ubiquity disfavors the ejection of a thin Ca-rich filament
that would only be observed from restricted angles. Moreover,
expanding shells are likely to wrap around narrow filaments and thus
we would not expect the relation between the \ion{Si}{2} cutoff and
\ion{Ca}{2} HV. Nonetheless, \ion{Ca}{2} polarization observed in
SN~2001el \citep{wang2003,kasen2003} requires some sort of asymmetry,
and consequently the mass estimates for the shell containing the
\ion{Ca}{2} HV are upper limits which may need to be reduced by
factors of 2 to 3. Recently, \citet{mazzali2005b} suggested that the
presence of high velocity Si is evidence for deflagrations, which we
find unlikely: \ion{Mg}{2} in the IR excludes mixing of a deflagration
model (see \S{\ref{SiII}}) and the ubiquity of \ion{Si}{2} at high
velocities disfavors line of sight effects. Moreover the relation of
the \ion{Si}{2} cutoff to the \ion{Ca}{2} HV points to a common
origin.
 
In our sample of early time spectra, most of the SNe~Ia can be
understood within the framework of the same model; however, there is a
remarkable exception to the homogeneity with respect to the light
curves, early time spectra, and evolution of the velocities: SN~1990N
and the similar event SN~2001el \citep{mattila2005}. This may be
regarded as a hint of distinct SNe~Ia sub-classes and lend support to
earlier suggestions attributing this division to different progenitors
such as pulsation delayed detonations or mergers
\citep{hoflich_khokhlov1996}.

\acknowledgments We would like to thank the staff of the Hobby-Eberly
Telescope and McDonald Observatory for their support, the HESS site
staff, and the ROTSE collaboration. This research is supported, in
part, by NASA grant NAG 5-7937 (PH) and NSF grants AST0307312 (PH) and
AST0406740 (RQ \& JCW). SJK is supported by an NSF Astronomy and
Astrophysics Postdoctoral Fellowship under award AST-0401547. CLG is
supported through UK PPARC grant PPA/G/S/2003/00040.

\end{document}